\documentclass[11pt,times,letterpaper]{article}
\usepackage{amsfonts,amssymb}
\usepackage{csquotes}

\usepackage{color}
\usepackage{multirow}
\usepackage{amsmath}
\usepackage{url}
\usepackage{graphicx}
\usepackage{ntheorem}
 \usepackage{mathrsfs}
\usepackage{cite}
\usepackage{enumitem}
 \usepackage{cases}
\newif\ifdraft
\draftfalse % or \draftfalse

 \newfont{\secfnta}{ptmb8t at 12pt}

\usepackage{makecell}
\usepackage[margin=.6in]{geometry}
\usepackage{relsize}

\newcommand{\1}[1]{{\bf 1}\left[#1\right]}

\newcommand{\bcap} {\hspace{2pt} \mathlarger{\cap}
\hspace{2pt}}

\newcommand{\ccap} {\hspace{2pt} \cap
\hspace{2pt}}

\newcommand{\bcup} {\hspace{2pt} \mathlarger{\cup}
\hspace{2pt}}
    \makeatletter
    \def\@listi{\leftmargin\leftmargini
        \parsep 1\p@ \@plus0\p@ \@minus\p@
        \topsep 2\p@   \@plus0\p@ \@minus\p@
        \itemsep1\p@ \@plus0\p@ \@minus\p@}
    \let\@listI\@listi\@listi
    \makeatother

\newcommand*\mcapinn[2]{\vcenter{\hbox{$\mathsurround=0pt
\ifx\displaystyle#1\textstyle\else#1\fi\bigcap$}}}

\newcommand*\mcupinn[2]{\vcenter{\hbox{$
\bigcup$}}}

\newcommand{\bP}[1]{{\mathbb{P}}\left[{#1}\right]}
\newcommand{\bE}[1]{{\mathbb{E}}\left[{#1}\right]}

\newtheorem{lem}{Lemma}
\newtheorem{thm}{Theorem}

\newtheorem{cor}{Corollary}
\newtheorem{proposition}{Proposition}

\begin{document}

\title{On connectivity in a general random intersection graph}

\author{Jun Zhao\\~\\Dept.
of ECE \\
Carnegie Mellon University \\
junzhao@alumni.cmu.edu}

\date{August 15, 2015}

\maketitle

%\author{{Jun Zhao}
%CyLab and Dept.
%of ECE \\
%Carnegie Mellon University \\
%Pittsburgh, PA 15213\\
%Email: junzhao@cmu.edu \and {Osman Ya\u{g}an}
%CyLab and Dept.
%of ECE\\
%Carnegie Mellon University \\
%Moffett Field, CA 94035\\
%Email: oyagan@ece.cmu.edu \and {Virgil Gligor}
%CyLab and Dept.
%of ECE \\
%Carnegie Mellon University \\
%Pittsburgh, PA 15213\\
%Email: gligor@cmu.edu}

%
%\author{\IEEEauthorblockN{Jun Zhao, Osman Ya\u{g}an and Virgil Gligor}
%\IEEEauthorblockA{\\CyLab and Dept.
%of ECE \\
%Carnegie Mellon University \\ \{junzhao, oyagan,
%virgil\}@andrew.cmu.edu}}

%\author{\IEEEauthorblockN{Jun Zhao, Osman Ya\u{g}an and Virgil Gligor}
%\IEEEauthorblockA{{\tiny~\vspace{-7pt}}\\CyLab and Dept.
%of ECE \\
%Carnegie Mellon University \\ \{junzhao, oyagan,
%virgil\}@andrew.cmu.edu\vspace{-7pt}}}

%\author{\IEEEauthorblockN{~}
%\IEEEauthorblockA{~\\
%~ \\ ~}}

%\markboth{} {\large This paper has been accepted at IEEE ISIT 2014
%to be held in June--July 2014.}

\begin{abstract}

%This paper, we investigate connectivity in a general random intersection graph.

There has been growing interest in studies of general random intersection graphs. In this paper, we consider
a general random intersection graph $\mathbb{G}(n,\overrightarrow{a}, \overrightarrow{K_n},P_n)$ defined on a set $\mathcal{V}_n$ comprising $n$ vertices, where $\overrightarrow{a}$ is a probability vector $(a_1,a_2,\ldots,a_m)$ and $\overrightarrow{K_n}$ is $(K_{1,n},K_{2,n},\ldots,K_{m,n})$. This graph     has been studied in the literature \cite{GodehardtJaworski,GodehardtJaworskiRybarczyk,yagan2015zero,ZhaoCDC} including a most recent work by Ya\u{g}an \cite{yagan2015zero}. Suppose  there is a pool $\mathcal{P}_n$ consisting of $P_n$ distinct objects. The $n$ vertices in $\mathcal{V}_n$ are divided into $m$ groups $\mathcal{A}_1, \mathcal{A}_2, \ldots, \mathcal{A}_m$. Each vertex $v$ is independently assigned to exactly a group according to the probability distribution with $\mathbb{P}[v \in \mathcal{A}_i]= a_i$, where $i=1,2,\ldots,m$. Afterwards, each vertex in group $\mathcal{A}_i$ independently chooses $K_{i,n}$ objects uniformly at random from the object pool $\mathcal{P}_n$. Finally, an undirected edge is drawn between two vertices in $\mathcal{V}_n$ that share at least one object. This graph model $\mathbb{G}(n,\overrightarrow{a}, \overrightarrow{K_n},P_n)$ has applications in secure sensor networks and social networks. We investigate connectivity in this general random intersection graph $\mathbb{G}(n,\overrightarrow{a}, \overrightarrow{K_n},P_n)$ and present a sharp zero-one law. Our result is also compared with the zero-one law established by Ya\u{g}an \cite{yagan2015zero}.

\iffalse

 For all vertices in

Each of the $n$
sensors in the network is classified into
$m$ groups according to some probability distribution $\overrightarrow{a}=\{a_1,\ldots,a_m\}$.
Before deployment, a group-$i$ sensor is assigned $K_i$ cryptographic keys that are selected uniformly at random from a common pool of $P$ keys.

Random

In this paper, we

  The $q$-composite key predistribution scheme \cite{adrian}
is used prevalently for secure communications in large-scale
wireless sensor networks (WSNs). Prior work \cite{yagan_onoff, ISIT,
ZhaoYaganGligor} explores secure connectivity in WSNs employing
the $q$-composite scheme for $q=1$ with unreliable communication
links modeled as independent on/off channels. In this paper, we
investigate secure connectivity in
WSNs operating under the $q$-composite scheme
for general $q$
 and under the on/off channel
model. We present conditions on how to scale the model parameters so that the network is securely connected with high probability when the number of sensors becomes large. The results are given in the form of  zero-one laws.

\fi

\end{abstract}

%\vspace{10pt}
%
%\begin{IEEEkeywords}
%key predistribution, vertex degree, random graph, random intersection
%graph, random key graph, security, topological properties, wireless
%sensor networks.
% \end{IEEEkeywords}


\section{Introduction}

Recently, there has been  considerable attention in analyzing random intersection graphs \cite{Rybarczyk,ZhaoYaganGligor,ZhaoCDC,yagan,ISIT,virgil,ZhaoAllerton,ball2014,mil10,2013arXiv1301.0466R,zz,r1,ryb3,bloznelis2013,virgillncs}. In a random intersection graph, each vertex is assigned to a set
of items in a random manner, and two vertices have an undirected edge in between if and only if they
have at least some number of items in common. In a specific model for a \emph{uniform} random intersection graph \cite{ryb3,r1,ZhaoCDC}, each vertex is independently assigned the same number of objects uniformly at random from a pool comprising different objects, and an undirected edge is drawn between two vertices that share at least one object. In the literature, the uniform random intersection graph model has been extensively studied \cite{Rybarczyk,ZhaoYaganGligor,yagan,ISIT,virgil,ZhaoAllerton,zz,r1,ryb3,bloznelis2013,virgillncs}, and there has been an increasing interest in investigating \emph{general} random intersection graphs \cite{GodehardtJaworski,GodehardtJaworskiRybarczyk,yagan2015zero,ZhaoCDC}.

In this paper, we look at
a general random intersection graph defined as below. We consider a graph defined on a set $\mathcal{V}_n$ with $n$ vertices. All vertices are divided into $m$ different groups $\mathcal{A}_1, \mathcal{A}_2, \ldots, \mathcal{A}_m$. Specifically, each vertex $v\in \mathcal{V}_n$ is independently assigned to exactly one group according to the following probability distribution\footnote{We summarize the notation and convention as follows. Throughout the paper, $\bP{\cdot}$ denotes a probability and $\bE{\cdot}$ stands for the expectation of a random variable. All limiting statements are understood with $n \to \infty$. We use the standard
asymptotic notation $o(\cdot), O(\cdot), \Omega(\cdot), \omega(\cdot),
\Theta(\cdot), \sim$. In
particular, for two positive sequences $f_n$ and $g_n$, the relation $f_n \sim
g_n$ means $\lim_{n \to
  \infty} ( {f_n}/{g_n})=1$. Also, ``$\ln $'' stands for the natural logarithm function, and ``$|\cdot|$'' can denote the absolute value as well as the
cardinality of a set.}: $\bP{v \in \mathcal{A}_i}= a_i$ for $i=1,2,\ldots,m$, where $m$ is a positive constant integer, and $a_i|_{i=1,2,\ldots,m}$ are positive constants satisfying the natural condition $\sum_{i=1}^{m}a_i  = 1$ (note that $m$ and $a_i|_{i=1,2,\ldots,m}$ do not scale with $n$). The edge set is built as follows. To begin with, assume that there exists a pool $\mathcal{P}_n$ consisting of $P_n$ distinct objects. Then for $i=1,2,\ldots,m$, each vertex in group $\mathcal{A}_i$ independently chooses $K_{i,n}$ objects uniformly at random from the object pool $\mathcal{P}_n$, where $1\leq K_{i,n}\leq P_n$. Finally, any two vertices in $\mathcal{V}_n$ have an undirected edge in between if and only if they share at least one object. With vectors $\overrightarrow{a}$ and $\overrightarrow{K_n}$ given by $\overrightarrow{a}=(a_1,a_2,\ldots,a_m)$ and $\overrightarrow{K_n}=(K_{1,n},K_{2,n},\ldots,K_{m,n})$, respectively, we denote the graph constructed above by $\mathbb{G}(n,\overrightarrow{a}, \overrightarrow{K_n},P_n)$. This graph has been investigated in the literature \cite{GodehardtJaworski,GodehardtJaworskiRybarczyk,yagan2015zero,ZhaoCDC}.
 For such a graph $\mathbb{G}(n,\overrightarrow{a}, \overrightarrow{K_n},P_n)$, we establish a zero-one law for connectivity:
\begin{displayquote}
~~~~~{\em For a graph $\mathbb{G}(n,\overrightarrow{a}, \overrightarrow{K_n},P_n)$ under $ P_n =
\Omega(n)$ and $\omega(1) = K_{1,n} \leq K_{2,n}\leq \ldots \leq K_{m,n} = o(\sqrt{P_n})$,}\\\indent~~~~~{\em if there exists a sequence $\beta_n$ satisfying $|\beta_n|= o(\ln n)$ such that
\begin{align}
\sum_{j=1}^{m} \left\{ a_j \left[1 -{{{P_n-K_{1,n}}\choose{K_{j,n}}}\over{{P_n}\choose{K_{j,n}}}}\right] \right\} & = \frac{\ln  n   +
 {\beta_n}}{n},   \label{eq:scalinglawxaa}
\end{align}}
~~~~~{\em then it holds that
\begin{subnumcases}
{ \hspace{-6pt}  \lim_{n \rightarrow \infty }\hspace{-1pt} \mathbb{P}\hspace{-1pt}\bigg[
\hspace{-3pt}\begin{array}{c}
\mathbb{G}(n,\overrightarrow{a}, \overrightarrow{K_n},P_n) \\
\mbox{is connected.}
\end{array}\hspace{-3pt}
\bigg] \hspace{-2pt}=\hspace{-2pt}}  \hspace{-3pt}0,\quad\hspace{-4pt}\text{if  }\lim_{n \to \infty}{\beta_n}  \hspace{-1.5pt} =  \hspace{-1.5pt} - \infty, \label{thm-con-eq-0xaa} \\
\hspace{-3pt}1,\quad\hspace{-4pt}\text{if  }\lim_{n \to \infty}{\beta_n}    \hspace{-1.5pt}=  \hspace{-1.5pt}  \infty. \label{thm-con-eq-1xaa}
\end{subnumcases}}
\end{displayquote}

For a general random intersection graph $\mathbb{G}(n,\overrightarrow{a}, \overrightarrow{K_n},P_n)$,
 (\ref{thm-con-eq-0xaa}) (resp., (\ref{thm-con-eq-1xaa})) present a zero-law (resp., one-law) for connectivity. This zero-one law  indicates that a \emph{critical} scaling for connectivity in a general random intersection graph $\mathbb{G}(n,\overrightarrow{a}, \overrightarrow{K_n},P_n)$ is to have the quantity in the left hand side of (\ref{eq:scalinglawxaa}) being $\frac{\ln n}{n}$, and $\beta_n$ in measures how much this quantity deviates from the critical value $\frac{\ln n}{n}$. Moreover, the zero-one law is \emph{sharp} since it suffices to have an unbounded deviation $\beta_n$ for (\ref{thm-con-eq-0xaa}) and (\ref{thm-con-eq-1xaa}); e.g., $\beta_n$ could be $\pm \Theta (\ln n \ln n), \pm \Theta (\ln n \ln n\ln n)$, etc. (We also note that our result has a condition $|\beta_n| = o(\ln n)$ for the proof to get through.)

 We explain below applications of our result to secure wireless sensor networks and social networks. In large-scale wireless sensor networks, an recognized approach to secure sensor communications is random key predistribution, where sensors  are equipped with cryptographic keys before deployment and uses the shared keys to establish secure communication after deployment. Among various random key predistribution schemes, a scheme proposed by Eschenauer and Gligor (EG) \cite{virgil} has gained the most attention. In the EG scheme, the memory of each sensor before deployment has a number of cryptographic keys selected uniformly at random from a common pool comprising distinct keys, and two sensors are able to communicate securely if they share at least one key. As explained below, a general random intersection graph $\mathbb{G}(n,\overrightarrow{a}, \overrightarrow{K_n},P_n)$ represents the topology of a secure sensor network employing a variation of the EG scheme in consideration of sensor heterogeneity, with the notion of an object in  the graph construction is specified as a cryptographic key. In a secure sensor network, all $n$ sensors are classified into $m$ groups $\mathcal{A}_1, \mathcal{A}_2, \ldots, \mathcal{A}_m$. Sensors in distinct groups may have different memory resources and thus are equipped with different number of keys. Specifically, for $i=1,2,\ldots,m$, each sensor in group $\mathcal{A}_i$ independently selects $K_{i,n}$ keys uniformly at random from a pool $\mathcal{P}_n$ comprising $P_n$ different keys. Clearly, our  result provides an analytical guideline on how to choose network parameters so that the secure sensor networks is connected with high probability. We further explain that the conditions
 $ P_n =
\Omega(n)$ and $\omega(1) = K_{1,n} \leq K_{2,n}\leq \ldots \leq K_{m,n} = o(\sqrt{P_n})$ are practical in secure sensor networks. First, $K_{1,n} \leq K_{2,n}\leq \ldots \leq K_{m,n}$ are assumed without loss of generality. Second, the key pool size $P_n$ grows at least linearly with the number of sensors $n$ and the number of keys on a sensor increases with $n$ becomes larger  to have reasonable resiliency against sensor capture attacks \cite{DiPietroTissec,yagan_onoff,ZhaoYaganGligor}, so $ P_n =
\Omega(n)$ are $K_{1,n} = \omega(1)$ both practical. Finally, since the number of keys on a sensor is often bounded above by a polylogarithmic function of $n$ since sensors have limited memory to store keys \cite{DiPietroTissec,yagan_onoff,ZhaoYaganGligor} and $P_n$ is  $\Omega(n)$ as mentioned above, $K_{m,n} = o(\sqrt{P_n})$ is also applicable to practical sensor networks.

A general random intersection graph $\mathbb{G}(n,\overrightarrow{a}, \overrightarrow{K_n},P_n)$ can also be used to model a social network, where a vertex represents an individual, and an object could be an hobby of individuals, a book being read, or a movie being watched, etc. Then a link between two people characterize a common-interest relation \cite{Assortativity,bloznelis2013,mil10,ZhaoYaganGligor,r4}; namely, two users have a connection if they have a common hobby, read a common book, or watch a common movie, etc. The heterogeneity of groups in $\mathbb{G}(n,\overrightarrow{a}, \overrightarrow{K_n},P_n)$ takes into account of the fact that users may have different number of interests. Our result   shed light on the effect of heterogeneity on   connectivity of a common-interest social network.

We organize the rest of the paper as follows. Section
2 presents some preliminaries.
Afterwards, we detail the main result in Section
\ref{sec:res}. Subsequently, Sections 4 and 5 are devoted to proving the results. %Section \ref{sec:disc} presents a discussion and
 Section \ref{related} surveys related work. Finally,  we conclude the paper in Section \ref{sec:Conclusion}.

 \section{Preliminaries}
\label{sec:Preliminaries}

We notate the $n$ vertices in graph $\mathbb{G}(n,\overrightarrow{a}, \overrightarrow{K_n},P_n)$ by $v_1,
v_2, \ldots, v_n$; i.e., $\mathcal {V} = \{v_1,
v_2, \ldots, v_n \}$.  For each
$x=1,2\ldots,n$, the object set of vertex $v_x$ is denoted by $S_x$. When $v_x$ belongs to a group $\mathcal{A}_i$ for some $i \in \{1,2,\ldots,m\}$, the set $S_x$ is uniformly distributed among all
$K_{i,n}$-size subsets of the object pool $\mathcal{P}_n$.

In graph $\mathbb{G}(n,\overrightarrow{a}, \overrightarrow{K_n},P_n)$, let $E_{xy}$ be the event that two different vertices $v_x$ and $v_y$ have an edge in between. Clearly, $E_{xy}$ is equivalent to the event $S_{x} \cap S_{y} \neq \emptyset$. To analyze $E_{xy}$, we often condition on the case where $v_x$ belongs to group $\mathcal{A}_i$ and $v_y$ belongs to group $\mathcal{A}_j$, where $i \in \{1,2,\ldots,m\}$ and $j \in \{1,2,\ldots,m\}$ (note that $x$ and $y$ are different, but $i$ and $j$ may be the same; i.e., different vertices $v_x$ and $v_y$ may belong to the same group).

Under $(v_x\in\mathcal{A}_i)\ccap (v_y\in\mathcal{A}_j)$, vertex $v_x$ has $K_{i,n}$ objects and vertex $v_y$ has $K_{j,n}$ objects. Then it is clear that $\mathbb{P} [E_{xy}~|~(v_x\in\mathcal{A}_i)\ccap (v_y\in\mathcal{A}_j) ]$ depends on $i$ and $j$, but does not rely on $x$ and $y$, so we can define
\begin{align}
p_{ij} & =  \mathbb{P} [E_{xy}~|~(v_x\in\mathcal{A}_i)\ccap (v_y\in\mathcal{A}_j) ]. \label{psq1contaxad}
\end{align}
We compute the probability $p_{ij}$ below. Let $T(K_{i,n},P_n)$ be the set of all $K_{i,n}$-size subsets of the object pool $\mathcal{P}_n$. Under $(v_x\in\mathcal{A}_i)\ccap (v_y\in\mathcal{A}_j)$, the set $S_x$ (resp., $S_y$) is uniformly distributed in $T(K_{i,n},P_n)$ (resp., $T(K_{j,n},P_n)$). Let $S_x^*$ be an arbitrary element in $T(K_{i,n},P_n)$. Conditioning on $S_x=S_x^*$, the event $\overline{E_{xy}}$ (i.e., $S_{x} \cap S_{y} = \emptyset$) means $S_{y}\subseteq \mathcal{P}_n \setminus S_x^*$. Noting that there are ${P_n}\choose{K_{j,n}}$ ways to select a $K_{j,n}$-size set from $\mathcal{P}_n$ and there are ${P_n-K_{i,n}}\choose{K_{j,n}}$ ways to select a $K_{j,n}$-size set from $\mathcal{P}_n\setminus S_x^*$, we readily obtain
\begin{align}
\bP{\overline{E_{xy}}~|~(S_x=S_x^*)\ccap (v_y\in\mathcal{A}_j)} & = {{{P_n-K_{i,n}}\choose{K_{j,n}}}\over{{P_n}\choose{K_{j,n}}}}. \label{psq1conta7sterwe}
\end{align}
From (\ref{psq1contaxad}) and (\ref{psq1conta7sterwe}), it follows that
 \begin{align}
p_{ij} & = \sum_{S_x^* \in T(K_{i,n},P_n)} \big\{ \bP{S_x=S_x^*} \bP{{E_{xy}}~|~(S_x=S_x^*)\ccap (v_y\in\mathcal{A}_j)} \big\} = 1- {{{P_n-K_{i,n}}\choose{K_{j,n}}}\over{{P_n}\choose{K_{j,n}}}}, \label{psq1contaxadrst}
\end{align}
where we use $\sum_{S_x^* \in T(K_{i,n},P_n)} \bP{S_x=S_x^*~|~ v_x\in \mathcal{A}_i}=1$.

Then we compute  $b_{i,n}$ which denotes $\mathbb{P} [E_{xy}~|~(v_x\in\mathcal{A}_i)  ]$; i.e., the probability of $v_x$ and $v_y$ have an edge conditioning on the event that $v_x$ belongs to group $\mathcal{A}_i$. Clearly, we have
\begin{align}
b_{i,n} & =\mathbb{P} [E_{xy}~|~(v_x\in\mathcal{A}_i)  ] \nonumber \\ & = \sum_{j=1}^{m} \big( \bP{v_y\in\mathcal{A}_j} \mathbb{P} [E_{xy}~|~(v_x\in\mathcal{A}_i)\ccap (v_y\in\mathcal{A}_j) ] \big) \nonumber \\ & = \sum_{j=1}^{m} \big( a_j p_{ij}\big). \label{psq1conta7ttt}
\end{align}

We can further compute $ \mathbb{P} [E_{xy} ]$. It follows that
\begin{align}
 \mathbb{P} [E_{xy} ]  & = \sum_{i=1}^{m} \big( \bP{v_x\in\mathcal{A}_i} \mathbb{P} [E_{xy}~|~(v_x\in\mathcal{A}_i)  ] \big) \nonumber \\ & = \sum_{i=1}^{m} \sum_{j=1}^{m} \big(a_i a_j p_{ij}\big). \label{psq1conta7tttaa}
\end{align}
where the last step uses (\ref{psq1conta7ttt}).

Our main result provided in the next section involves $b_{1,n}$ (see (\ref{eq:scalinglaw})), which is the probability that a typical vertex in group $\mathcal{A}_1$ has an edge with another typical vertex in $\mathcal{V}_n$. From (\ref{psq1contaxadrst}) and (\ref{psq1conta7ttt}) with $i=1$, we obtain that $b_{1,n}$ equals the left hand side of (\ref{eq:scalinglawxaa}).

\section{The Main Result} \label{sec:res}

We present the main result in Theorem 1 below.

\iffalse
%$\mathbb{N}_0 $ stands for the set of all positive integers;
%$\mathbb{R}$ is the set of all real numbers;
The
natural logarithm function is given by $\ln$. All limits are understood with $n
\to
  \infty$.  %The term ``for all $n$
%sufficiently large'' means ``for any $n \geq N$, where $N \in
%\mathbb{N}_0$ is selected appropriately''.
 We use the standard
asymptotic notation $o(\cdot), O(\cdot), \Omega(\cdot), \omega(\cdot),
\Theta(\cdot), \sim$.

Theorem \ref{thm:OneLaw+NodeIsolation} below presents

\fi

\begin{thm}  \label{thm:OneLaw+NodeIsolation}

Consider a general random intersection
graph $\mathbb{G}(n,\overrightarrow{a}, \overrightarrow{K_n},P_n)$ under $ P_n =
\Omega(n)$ and $\omega(1) = K_{1,n} \leq K_{2,n}\leq \ldots \leq K_{m,n} = o(\sqrt{P_n})$.
%$K_{m,n} \leq \sqrt{P_n n^{\varepsilon-1}}$ for all $n$ sufficiently large, where $\varepsilon$ is an arbitrary positive constant.
 If there exists a sequence $\beta_n$ satisfying $|\beta_n|= o(\ln n)$ such that
\begin{align}
b_{1,n}  & = \frac{\ln  n   +
 {\beta_n}}{n},   \label{eq:scalinglaw}
\end{align}
where $b_{1,n}$ equals the left hand side of (\ref{eq:scalinglawxaa}), then it holds that
\begin{subnumcases}
{ \hspace{-6pt}  \lim_{n \rightarrow \infty }\hspace{-1pt} \mathbb{P}\hspace{-1pt}\bigg[
\hspace{-3pt}\begin{array}{c}
\mathbb{G}(n,\overrightarrow{a}, \overrightarrow{K_n},P_n) \\
\mbox{is connected.}
\end{array}\hspace{-3pt}
\bigg] \hspace{-2pt}=\hspace{-2pt}}  \hspace{-3pt}0,\quad\hspace{-4pt}\text{if  }\lim_{n \to \infty}{\beta_n}  \hspace{-1.5pt} =  \hspace{-1.5pt} - \infty, \label{thm-con-eq-0} \\
\hspace{-3pt}1,\quad\hspace{-4pt}\text{if  }\lim_{n \to \infty}{\beta_n}    \hspace{-1.5pt}=  \hspace{-1.5pt}  \infty. \label{thm-con-eq-1}
\end{subnumcases}
\end{thm}

In Section I, we have already discussed the result in Theorem \ref{thm:OneLaw+NodeIsolation}; in particular, we explain therein that Theorem \ref{thm:OneLaw+NodeIsolation} gives  a sharp zero-one law for connectivity in a  graph $\mathbb{G}(n,\overrightarrow{a}, \overrightarrow{K_n},P_n)$. The next section presents the proof of Theorem \ref{thm:OneLaw+NodeIsolation}.

\section{Establishing Theorem 1}
%\ref{thm:OneLaw+NodeIsolation}}
\label{sec:ProofTheoremNodeIsolation}

In proving Theorem \ref{thm:OneLaw+NodeIsolation}, we use the relationship between connectivity and the absence of isolated vertex. It is easy to see that if a graph is connected, then
it does not contain any~isolated~vertex \cite{Gupta98criticalpower}. Therefore,  we immediately have
\begin{align}
 \mathbb{P} \left[\hspace{2pt}\mathbb{G}(n,\overrightarrow{a},\overrightarrow{K_n},P_n)\text{
is connected}.\hspace{2pt}\right]
\leq  \bP{
\hspace{-2pt}\begin{array}{c}
\mathbb{G}(n,\overrightarrow{a},\overrightarrow{K_n},P_n)~\mbox{contains} \\
\mbox{~no~isolated~vertex.}
\end{array}\hspace{-2pt} \label{pcond1}
}
 \end{align}
and
\begin{align}
& \mathbb{P} \left[\hspace{2pt}\mathbb{G}(n,\overrightarrow{a},\overrightarrow{K_n},P_n)\text{
is connected}.\hspace{2pt}\right]
\nonumber \\ & \quad =  \bP{
\hspace{-2pt}\begin{array}{c}
\mathbb{G}(n,\overrightarrow{a},\overrightarrow{K_n},P_n)~\mbox{contains} \\
\mbox{~no~isolated~vertex.}
\end{array}\hspace{-2pt}
} - \mathbb{P} \bigg[\hspace{-3pt}\begin{array}{l}
\mathbb{G}(n,\overrightarrow{a},\overrightarrow{K_n},P_n)\text{
has no isolated vertex}, \\\text{but is not connected.}
\end{array}
\hspace{-3pt}\bigg]. \label{pcond2}
 \end{align}

 Given (\ref{pcond1}) and  (\ref{pcond2}), we will complete proving Theorem \ref{thm:OneLaw+NodeIsolation} once we establish  Lemmas \ref{lem_Gqsbsd} and \ref{lem_Gq_no_isolated_but_not_conn} below. In the rest of the paper, by ``the conditions of Theorem \ref{thm:OneLaw+NodeIsolation}'', we mean $ P_n =
\Omega(n)$, $\omega(1) = K_{1,n} \leq K_{2,n}\leq \ldots \leq K_{m,n} = o(\sqrt{P_n})$ and (\ref{eq:scalinglaw}) (i.e., $b_{1,n}  = \frac{\ln  n   +
 {\beta_n}}{n}$) with $|\beta_n| = o(\ln n) $.

\begin{lem} \label{lem_Gqsbsd}

{
For a graph $\mathbb{G}(n,\overrightarrow{a}, \overrightarrow{K_n},P_n)$ under the conditions of Theorem \ref{thm:OneLaw+NodeIsolation}, it holds that
\begin{subnumcases}
{ \hspace{-6pt}  \lim_{n \rightarrow \infty }\hspace{-1pt} \mathbb{P}\hspace{-1pt}\bigg[
\hspace{-3pt}\begin{array}{c}
\mathbb{G}(n,\overrightarrow{a},\overrightarrow{K_n},P_n)~\mbox{contains} \\
\mbox{~no~isolated~vertex.}
\end{array}\hspace{-3pt}
\bigg] \hspace{-2pt}=\hspace{-2pt}}  \hspace{-3pt}0,\quad\hspace{-4pt}\text{if  }\lim_{n \to \infty}{\beta_n}  \hspace{-1.5pt} =  \hspace{-1.5pt} - \infty, \label{thm-mnd-eq-0} \\
\hspace{-3pt}1,\quad\hspace{-4pt}\text{if  }\lim_{n \to \infty}{\beta_n}    \hspace{-1.5pt}=  \hspace{-1.5pt}  \infty, \label{thm-mnd-eq-1}
\end{subnumcases}
 }
\end{lem}

Lemma \ref{lem_Gqsbsd} presents a zero-one law on the absence of isolated vertex via  (\ref{thm-mnd-eq-0}) and (\ref{thm-mnd-eq-1}). In the next subsection, we explain the idea of proving (\ref{thm-mnd-eq-0}) and (\ref{thm-mnd-eq-1}) by the method of moments.

\begin{lem} \label{lem_Gq_no_isolated_but_not_conn}

{
For a graph $\mathbb{G}(n,\overrightarrow{a}, \overrightarrow{K_n},P_n)$ under the conditions of Theorem \ref{thm:OneLaw+NodeIsolation}, it holds that
\begin{align}
\lim_{n \to \infty}\mathbb{P} \bigg[\hspace{-3pt}\begin{array}{c}
\mathbb{G}(n,\overrightarrow{a},\overrightarrow{K_n},P_n)\text{
has no isolated vertex}, \\\text{but is not connected.}
\end{array}
\hspace{-3pt}\bigg] = 0.  \label{eq:OneLawAfterReductionsb}
 \end{align}
 }
\end{lem}

Lemma \ref{lem_Gq_no_isolated_but_not_conn} is established in Section \ref{sec:lem_Gq_no_isolated_but_not_conn}.

 % In proving Lemmas \ref{lem_Gqsbsd} and \ref{lem_Gq_no_isolated_but_not_conn}, we will use the following lemma.

%The proof of Lemma \ref{lemboundKn} in the Appendix.

\subsection{Method of moments to prove Lemma \ref{lem_Gqsbsd} on the absence of isolated vertex}

%$\mathcal{A}_1$

We use the method of moments \cite[Page 55]{JansonLuczakRucinski7} to prove Lemma \ref{lem_Gqsbsd} on the absence of isolated vertex. Below we establish (\ref{thm-mnd-eq-0}) and (\ref{thm-mnd-eq-1}) of Lemma \ref{lem_Gqsbsd}, respectively.

\subsubsection{Establishing (\ref{thm-mnd-eq-0})}

 We prove (\ref{thm-mnd-eq-0}) by the method of the first moment \cite[Page 55]{JansonLuczakRucinski7}
applied to the total number of isolated vertices in $\mathbb{G}(n,\overrightarrow{a},\overrightarrow{K_n},P_n)$.  With indicator variables $\phi_{n,i}$ for $i=1, \ldots , n$ defined by
\begin{align}
\phi_{n,i} & = \1{ {\rm Vertex~}v_i~\text{is~isolated~in~}
                         \mathbb{G}(n,\overrightarrow{a},\overrightarrow{K_n},P_n). } \nonumber\\
                         & =
\begin{cases}
1, &\text{if $v_i$ is isolated in $\mathbb{G}(n,\overrightarrow{a},\overrightarrow{K_n},P_n)$,}\\
0, &\text{if $v_i$ is not isolated in $\mathbb{G}(n,\overrightarrow{a},\overrightarrow{K_n},P_n)$.}
\end{cases}  \nonumber
\end{align}
then ${J_n}$ denoting the number of isolated vertex in
$\mathbb{G}(n,\overrightarrow{a},\overrightarrow{K_n},P_n)$ is given by
\[
{J_n}  := \sum_{i=1}^n \phi_{n,i}.
\]
The random graph $\mathbb{G}(n,\overrightarrow{a},\overrightarrow{K_n},P_n)$ has no isolated
vertex if and only if ${J_n}  = 0$.

The method of first moment \cite[Equation (3.10), Page
55]{JansonLuczakRucinski7} relies on the well-known bound
\begin{equation}
1 - \bE{ {J_n}  } \leq \bP{  {J_n}  = 0 }.
\label{eq:FirstMoment-one}
\end{equation}
Noting that the random variables $\phi_{n,1}, \ldots , \phi_{n,n} $ are
exchangeable due to vertex symmetry, we find
\begin{equation}
\bE{ {J_n} } = n \bE{ \phi_{n,1}  }
\label{eq:FirstMomentExpression-one}
\end{equation}

The desired one-law (\ref{thm-mnd-eq-1}) means $\lim_{n\to \infty} \bP{{J_n} = 0}
= 1$ under $\lim_{n \to \infty}{\beta_n}  =  \infty$. From (\ref{eq:FirstMoment-one}) and (\ref{eq:FirstMomentExpression-one}), $\lim_{n\to \infty} \bP{{J_n} = 0}
= 1$ will be proved once we show
\begin{equation}
\lim_{n \to \infty} \big(n\bE{ \phi_{n,1}  }\hspace{-.5pt}\big)= 0.
\label{eq:OneLaw+NodeIsolation+convergence-one}
\end{equation}

Clearly, the event $(\phi_{n,1} =1)$ (i.e.,
vertex $v_1$ is isolated) is equivalent to $\bigcap_{j=2}^{n}\overline{E_{1j}} $, where the event $\overline{E_{1j}}$ means that there is no edge between $v_1$ and $v_j$ in graph $\mathbb{G}(n,\overrightarrow{a}, \overrightarrow{K_n},P_n)$. %and is equivalent to $  \overline{B_{1j}} \bcup \overline{\Gamma_{1j}}$, with
%$\overline{\Gamma_{1j}}$ being $(|S_1 \cap S_j| <
%q)$.
 %For each $j =2,\ldots,n$, event $\overline{B_{1j}}$ is independent of events $\overline{\Gamma_{1j}}$,  $\overline{B_{1j'}}$ and $\overline{\Gamma_{1j'}}$ for $j'\in\{2,\ldots,n\}\setminus\{j\}$.
 % Conditioning on $S_1$, events $\overline{B_{1j}}|_{j=1,\ldots,n}$ are independent. Hence, from $\overline{E_{1j}} =  \overline{B_{1j}} \bcup \overline{\Gamma_{1j}}$, we know that
 %  Clearly, conditioning on $S_1$, the probability of $\overline{E_{1j}}|_{j=1,\ldots,n}$ is given by $1-s_n p_n = 1- t_n$. In view of the above, conditioning on $S_1$, the probability of $(\phi_{n,1} =1)$ is $\left(1-t_n\right)^{n-1}$.
    Then
\begin{align}
\bE{ \phi_{n,1}  } &=
 \bP{ v_1~\text{is~isolated} }
\nonumber \\
&=  \sum_{i=1}^{m} \bP{v_1\in \mathcal{A}_i }  \bP{ v_1~\text{is~isolated} ~ |~ v_1\in \mathcal{A}_i }
\nonumber \\
&= \sum_{i=1}^{m} a_i \bP{ \cap_{j=2}^{n} \overline{E_{1j}}  ~ |~ v_1\in \mathcal{A}_i }
%\nonumber \\
%&= \sum_{i=1}^{m} a_i \left(\bP{\overline{E_{12}}  ~ |~ v_1\in \mathcal{A}_i }\right)^{n-1}
\label{eq:intermed_one_law}
%\\
%&=& n \sum_{i=1}^{m} a_i \left(\bE{1 -p_{1c(2)}  }\right)^{n-1}
\end{align}
Recall that $T(K_{i,n},P_n)$ is the set of all $K_{i,n}$-size subsets of the object pool $\mathcal{P}_n$. Conditioning on $S_1$, events $\overline{E_{1j}}|_{j=2,\ldots,n}$ are independent. Moreover, given any $S_1^* \in T(K_{i,n},P_n)$, we note
\begin{align}
\bP{\overline{E_{1j}}  ~ |~ S_1=S_1^* } &=
1- \bP{ {E_{1j}}  ~ |~ S_1=S_1^* }
 = 1- b_{i,n},\quad \text{ for $j=2,\ldots,n$.}
\end{align}
where $b_{i,n}$ is given by (\ref{psq1conta7ttt}). Hence, it holds that
\begin{align}
\bP{ \cap_{j=2}^{n} \overline{E_{1j}}  ~ |~ v_1\in \mathcal{A}_i } &=
\sum_{S_1^* \in T(K_{i,n},P_n)} \bP{S_1=S_1^*}  \left(1- b_{i,n}\right)^{n-1}
 =  \left(1- b_{i,n}\right)^{n-1}, \label{eq:intermed_one_laweq1}
\end{align}
where the last step uses $\sum_{S_1^* \in T(K_{i,n},P_n)} \bP{S_1=S_1^*~|~ v_1\in \mathcal{A}_i}=1$. Then the application of (\ref{eq:intermed_one_laweq1}) to (\ref{eq:intermed_one_law}) yields
%\begin{align}
%\bE{ \phi_{n,1}  } &=
%\sum_{i=1}^{m} a_i \left(\bP{\overline{E_{12}}  ~ |~ v_1\in \mathcal{A}_i }\right)^{n-1}
%\nonumber \\
%&=
%\end{align}
%
%
%Then using $\bP{\overline{E_{12}}  ~ |~ v_1\in \mathcal{A}_i } =1- b_{i,n} $ from ??, we obtain
\begin{align}
 n \bE{ \phi_{n,1}  }
&= n \sum_{i=1}^{m} a_i (1- b_{i,n})^{n-1}
 \leq n (1- b_{1,n})^{n-1}  . \label{pntlnnalpanaa}
\end{align}

\iffalse
 we get by independence that
\begin{align}
\bE{ \psi_{n,1}  } & = \bP{\hspace{1.5pt}\bigcap_{j=2}^{n} \big(\hspace{1.5pt}\overline{B_{1j}} \bcup \overline{\Gamma_{1j}}\hspace{1.5pt}\big)} \nonumber \\
& =  \sum_{S_1^{*} \in} \bP{\hspace{1.5pt}\bigcap_{j=2}^{n} \big(\hspace{1.5pt}\overline{B_{1j}} \bcup \overline{\Gamma_{1j}}\hspace{1.5pt}\big)
~\bigg|~S_1=S_1^{*}} \bP{S_1=S_1^{*}}
 \nonumber \\
& =  \sum_{S_1^{*} \in} \bigg[ \hspace{1.5pt}\prod_{j=2}^{n} \bigg(\bP{\hspace{1.5pt}\overline{B_{1j}}\hspace{1.5pt}} \bP{\hspace{1.5pt} \overline{\Gamma_{1j}}
~\big|~S_1=S_1^{*}} \bigg) \bP{S_1=S_1^{*}} \bigg]
 \nonumber \\
& =
 \left(1-{p_n} s_n\right)^{n-1} =
\left(1-t_n\right)^{n-1}.
\end{align}
\fi

From (\ref{eq:scalinglaw}) and
$|\beta_n| = o(\ln n) $, it follows that
\begin{align}
b_{1,n} &   \sim   \frac{\ln  n}{n} = o(1) .  \label{pntlnnalp}
\end{align}
Then given $b_{1,n}  = o(1)$ and ${b_{1,n} }^2 \cdot (n-1)   \sim  \big( \frac{\ln  n}{n}
\big)^2 \cdot (n-1)= o(1)$,
we use \cite[Fact 3]{ZhaoYaganGligor} to derive
\begin{align}
& n (1- b_{1,n})^{n-1} \sim e^{ -b_{1,n}   (n-1)}.  \label{sbcc2}
\end{align}
 Substituting (\ref{eq:scalinglaw})   and $b_{1,n}  = o(1)$  into   (\ref{sbcc2}), we obtain
 \begin{align}
& n (1- b_{1,n})^{n-1}  \sim n  e^{ -n b_{1,n}}
\cdot e^{ b_{1,n}}
=n e^{-\ln n - \beta_n} \cdot e^{o(1)}
\sim e ^{- \beta_n}.  \label{sbcc2sbstrd}
\end{align}
 In view of (\ref{pntlnnalpanaa}) and  (\ref{sbcc2sbstrd}), we conclude
\begin{align}
&n \bE{ \psi_{n,1}  }  \leq e ^{- \beta_n} \cdot [1+o(1)] , \nonumber
\end{align}
which implies (\ref{eq:OneLaw+NodeIsolation+convergence-one}). Then as explained above, (\ref{thm-mnd-eq-0}) is proved.

\subsubsection{Establishing (\ref{thm-mnd-eq-1})}

We prove (\ref{thm-mnd-eq-1}) by the method of moments \cite[Page 55]{JansonLuczakRucinski7}
applied to the   number of  vertices that belong to group $\mathcal{A}_1$ and are isolated in $\mathbb{G}(n,\overrightarrow{a},\overrightarrow{K_n},P_n)$.  With indicator variables $\psi_{n,i}$ for $i=1, \ldots , n$ defined by
\begin{align}
\psi_{n,i} & = \1{ {\rm Vertex~}v_i~\text{belongs to group $\mathcal{A}_1$ and is~isolated~in~}
                         \mathbb{G}(n,\overrightarrow{a},\overrightarrow{K_n},P_n). } \nonumber\\
                         & =
\begin{cases}
1, &\text{if $v_i\in \mathcal{A}_1$ \textit{and} $v_i$ is isolated in $\mathbb{G}(n,\overrightarrow{a},\overrightarrow{K_n},P_n)$,}\\
0, &\text{if $v_i\notin \mathcal{A}_1$ \textit{or} $v_i$ is not isolated in $\mathbb{G}(n,\overrightarrow{a},\overrightarrow{K_n},P_n)$.}
\end{cases}  \nonumber
\end{align}
then ${I_n}$ denoting the number of isolated vertex in
$\mathbb{G}(n,\overrightarrow{a},\overrightarrow{K_n},P_n)$ is given by
\[
{I_n}  := \sum_{i=1}^n \psi_{n,i}.
\]
From the method of second moment \cite[Page
55]{JansonLuczakRucinski7}, it follows that
\begin{equation}
\bP{  {I_n}  = 0 } \leq 1 - \frac{ \bE{ {I_n} }^2}{
\bE{ {I_n}  ^2} }. \label{eq:SecondMoment}
\end{equation}
Noting that the random variables $\psi_{n,1}, \ldots , \psi_{n,n} $ are
exchangeable due to vertex symmetry, we find
\begin{equation}
\bE{ {I_n} } = n \bE{ \psi_{n,1}  }
\label{eq:FirstMomentExpression}
\end{equation}
and
\begin{align}
\bE{ {I_n} ^2 } & = n \bE{{\psi_{n,1}}^2   }
 +  n(n-1) \bE{ \psi_{n,1}   \psi_{n,2}  } \nonumber \\
 & = n \bE{ \psi_{n,1}  }
 +  n(n-1) \bE{ \psi_{n,1}   \psi_{n,2}  }, \label{eq:SecondMomentExpression}
\end{align}
where the last step uses $\bE{{\psi_{n,1}}^2   } = \bE{ \psi_{n,1}  }$
as $\psi_{n,1}$ is a binary random variable. It then follows from (\ref{eq:FirstMomentExpression}) and (\ref{eq:SecondMomentExpression}) that
\begin{eqnarray}
\frac{ \bE{ {I_n} ^2 }}{ \bE{ {I_n}  }^2 } = \frac{
1}{  n\bE{ \psi_{n,1}  } }
  + \frac{n-1}{n} \cdot \frac{\bE{ \psi_{n,1}
\psi_{n,2}  }}
     {\left (  \bE{ \psi_{n,1}  } \right )^2 }.
\label{eq:SecondMomentRatio}
\end{eqnarray}

 The desired zero-law (\ref{thm-mnd-eq-0}) means $\lim_{n\to \infty} \bP{{I_n} = 0}
= 0$ under $\lim_{n \to \infty}{\beta_n}  = - \infty$. From (\ref{eq:SecondMoment}) and
(\ref{eq:SecondMomentExpression}), we will obtain $\lim_{n\to \infty} \bP{{I_n} = 0}
= 0$ once deriving
\begin{equation}
\lim_{n \to \infty} \big(n \bE{ \psi_{n,1}  }\hspace{-.5pt}\big)= \infty
\label{eq:OneLaw+NodeIsolation+convergence2}
\end{equation}
and
\begin{equation}
 \frac{\bE{ \psi_{n,1}
\psi_{n,2}  }}
     {\left (  \bE{ \psi_{n,1}  }\hspace{-.5pt} \right )^2 }
 \leq 1 + o(1). \label{eq:ZeroLaw+NodeIsolation+convergence}
\end{equation}
The reason is that under (\ref{eq:OneLaw+NodeIsolation+convergence2}) and
(\ref{eq:ZeroLaw+NodeIsolation+convergence}), we apply them to (\ref{eq:SecondMomentRatio}) and derive
$\frac{ \bE{ {I_n} ^2 }}{ \bE{ {I_n}  }^2 }   \leq 1 + o(1)$, implying $\frac{ \bE{ {I_n}  }^2 }{ \bE{ {I_n} ^2 }}   \geq 1 -o(1)$, which is used in (\ref{eq:SecondMoment}) to establish $\lim_{n \to
\infty} \bP{{I_n} = 0} = 0$.

Below we prove (\ref{eq:OneLaw+NodeIsolation+convergence2}) and
(\ref{eq:ZeroLaw+NodeIsolation+convergence}), respectively.

\paragraph{Establishing (\ref{eq:OneLaw+NodeIsolation+convergence2}):}~

Clearly, the event $(\psi_{n,1} =1)$ is equivalent to $ (v_i\in \mathcal{A}_1)\bcap \big( \cap_{j=2}^{n}\overline{E_{1j}}\big)$. %and is equivalent to $  \overline{B_{1j}} \bcup \overline{\Gamma_{1j}}$, with
%$\overline{\Gamma_{1j}}$ being $(|S_1 \cap S_j| <
%q)$.
 %For each $j =2,\ldots,n$, event $\overline{B_{1j}}$ is independent of events $\overline{\Gamma_{1j}}$,  $\overline{B_{1j'}}$ and $\overline{\Gamma_{1j'}}$ for $j'\in\{2,\ldots,n\}\setminus\{j\}$.
 % Conditioning on $S_1$, events $\overline{B_{1j}}|_{j=1,\ldots,n}$ are independent. Hence, from $\overline{E_{1j}} =  \overline{B_{1j}} \bcup \overline{\Gamma_{1j}}$, we know that
 %  Clearly, conditioning on $S_1$, the probability of $\overline{E_{1j}}|_{j=1,\ldots,n}$ is given by $1-s_n p_n = 1- t_n$. In view of the above, conditioning on $S_1$, the probability of $(\psi_{n,1} =1)$ is $\left(1-t_n\right)^{n-1}$.
    Then
\begin{align}
\bE{ \psi_{n,1}  } &= \bP{\psi_{n,1} =1}  \nonumber \\
&=  \bP{(v_i\in \mathcal{A}_1)\bcap (\cap_{j=2}^{n}\overline{E_{1j}})} \nonumber \\
&=   \bP{v_i\in \mathcal{A}_1} \bP{ \cap_{j=2}^{n}\overline{E_{1j}}~ |~ v_1\in \mathcal{A}_1 }
\nonumber \\
&=  a_1 \bP{ \cap_{j=2}^{n} \overline{E_{1j}}  ~ |~ v_1\in \mathcal{A}_1 }
%\nonumber \\
%&= \sum_{i=1}^{m} a_i \left(\bP{\overline{E_{12}}  ~ |~ v_1\in \mathcal{A}_i }\right)^{n-1}
\label{eq:intermed_one_law-one}
%\\
%&=& n \sum_{i=1}^{m} a_i \left(\bE{1 -p_{1c(2)}  }\right)^{n-1}
\end{align}
From (\ref{eq:intermed_one_laweq1}), we have
\begin{align}
\bP{ \cap_{j=2}^{n} \overline{E_{1j}}  ~ |~ v_1\in \mathcal{A}_1 } &= \left(1- b_{1,n}\right)^{n-1}. \label{eq:intermed_one_laweq1-one}
\end{align}
  Then the use of (\ref{eq:intermed_one_laweq1-one}) to (\ref{eq:intermed_one_law-one}) induces
%\begin{align}
%\bE{ \psi_{n,1}  } &=
%\sum_{i=1}^{m} a_i \left(\bP{\overline{E_{12}}  ~ |~ v_1\in \mathcal{A}_i }\right)^{n-1}
%\nonumber \\
%&=
%\end{align}
%
%
%Then using $\bP{\overline{E_{12}}  ~ |~ v_1\in \mathcal{A}_i } =1- b_{i,n} $ from ??, we obtain
\begin{align}
  \bE{ \psi_{n,1}  }
&=   a_1 (1- b_{1,n})^{n-1}. \label{firstm}
\end{align}
Furthermore, from (\ref{sbcc2sbstrd}) and (\ref{firstm}), we derive
\begin{align}
 n \bE{ \psi_{n,1}  }
& \sim  a_1 e ^{- \beta_n}. \label{firstmnpx}
\end{align}
Since $a_1$ is a positive constant, (\ref{firstmnpx}) implies (\ref{eq:OneLaw+NodeIsolation+convergence2}).

\paragraph{Establishing (\ref{eq:ZeroLaw+NodeIsolation+convergence}):}~

The event $\big(\psi_{n,1}   =1\big) ~\cap~ \big( \psi_{n,2}  = 1\big)$ means that both vertices $v_1$ and $v_2$ belong to group $\mathcal{A}_1$ and are isolated in $\mathbb{G}(n,\overrightarrow{a},\overrightarrow{K_n},P_n)$. Then clearly event $\big(\psi_{n,1}   =1\big) ~\cap~ \big( \psi_{n,2}  = 1\big)$ is given by $\big[\cap_{j=3}^{n} \big(\hspace{1.5pt}  \overline{E_{1j}}
\cap  \overline{E_{2j}} \hspace{1.5pt} \big) \big]\cap (v_1\in \mathcal{A}_1) \cap (v_2\in \mathcal{A}_1) \cap  \overline{E_{12}}$. Therefore, we have
\begin{align}
 & \bE{ \psi_{n,1}  \psi_{n,2}  } \nonumber  \\
&  \quad  = \bP{\big(\psi_{n,1}   =1\big) ~\cap~ \big( \psi_{n,2}  = 1\big) } \nonumber  \\
& \quad = \bP{   \bigg[\bigcap_{j=3}^{n} \big(\hspace{1.5pt}  \overline{E_{1j}}
\cap  \overline{E_{2j}} \hspace{1.5pt} \big) \bigg]\cap (v_1\in \mathcal{A}_1) \cap (v_2\in \mathcal{A}_1) \cap  \overline{E_{12}} } \nonumber \\
& \quad = \bP{(v_1\in \mathcal{A}_1) \cap (v_2\in \mathcal{A}_1) \cap  \overline{E_{12}} } \bP{  \bigg[\bigcap_{j=3}^{n} \big(\hspace{1.5pt}  \overline{E_{1j}}
\cap  \overline{E_{2j}} \hspace{1.5pt} \big) \bigg]~|~ (v_1\in \mathcal{A}_1) \cap (v_2\in \mathcal{A}_1) \cap  \overline{E_{12}} }.\label{suba7}
\end{align}
First, we get
 \begin{align}
&  \bP{(v_1\in \mathcal{A}_1) \cap (v_2\in \mathcal{A}_1) \cap  \overline{E_{12}} } \nonumber \\
& \quad \leq
 \bP{(v_1\in \mathcal{A}_1) \cap (v_2\in \mathcal{A}_1) } \nonumber \\
& \quad =   \bP{v_1\in \mathcal{A}_1} \bP{v_2\in \mathcal{A}_1}  \nonumber \\
& \quad =  {a_1}^2. \label{suba6}
\end{align}
Second, to evaluate $ \bP{  [\bigcap_{j=3}^{n} \big(\hspace{1.5pt}  \overline{E_{1j}}
\cap  \overline{E_{2j}} \hspace{1.5pt} \big) ]~|~ (v_1\in \mathcal{A}_1) \cap (v_2\in \mathcal{A}_1) \cap  \overline{E_{12}} }$, we find it useful to define a set $\mathcal{L}(K_{1,n},P_n)$ as follows:
\begin{align}
\mathcal{L}(K_{1,n},P_n)= \big\{(L_1,L_2)~|~\big(L_1\in T(K_{1,n},P_n)\big)\cap\big(L_2\in T(K_{1,n},P_n)\big)\cap\big(L_1\cap L_2 = \emptyset\big)\big\}.
\end{align}
Then the event $(v_1\in \mathcal{A}_1) \cap (v_2\in \mathcal{A}_1) \cap  \overline{E_{12}}$ is equivalent to the event that the vector $(S_1,S_2)$ belongs to $\mathcal{L}(K_{1,n},P_n)$. Conditioning on $S_1$ and $S_2$, events $(\overline{E_{1j}}
\cap  \overline{E_{2j}})|_{j=3,\ldots,n}$ are independent. Hence, it becomes clear that
\begin{align}
& \bP{  \bigg[\bigcap_{j=3}^{n} \big(\hspace{1.5pt}  \overline{E_{1j}}
\cap  \overline{E_{2j}} \hspace{1.5pt} \big) \bigg]~|~ (v_1\in \mathcal{A}_1) \cap (v_2\in \mathcal{A}_1) \cap  \overline{E_{12}} } \nonumber \\
& \quad =
\sum_{(S_1^*,S_2^*) \in \mathcal{L}(K_{1,n},P_n)} \bigg\{\bP{(S_1=S_1^*) \cap (S_2=S_2^*)} \prod_{j=3}^{n} \bP{\overline{E_{1j}}
\cap  \overline{E_{2j}} ~ |~ (S_1=S_1^*) \cap (S_2=S_2^*) }\bigg\}. \label{suba1}
\end{align}
As the above summation shows, $(S_1^*,S_2^*)$ is an arbitrary element in $\mathcal{L}(K_{1,n},P_n)$.

For $j=3,\ldots,n$, we compute conditional probabilities when vertex $v_j$ falls into a group $\mathcal{A}_{\ell}$ for some $\ell$ in $\{1,\ldots,n\}$, so it follows that
\begin{align}
& \bP{\overline{E_{1j}}
\cap  \overline{E_{2j}} ~ |~ (S_1=S_1^*) \cap (S_2=S_2^*) }\nonumber \\
& \quad =
\sum_{\ell=1}^{m}\Big\{\bP{v_j \in \mathcal{A}_{\ell} }\bP{\overline{E_{1j}}
\cap  \overline{E_{2j}} ~ |~ (S_1=S_1^*) \cap (S_2=S_2^*)\cap (v_j \in \mathcal{A}_{\ell}) } \Big\}
\nonumber \\
& \quad =
\sum_{\ell=1}^{m}\Big\{a_{\ell}\hspace{1.5pt}\bP{\overline{E_{1j}}
\cap  \overline{E_{2j}} ~ |~ (S_1=S_1^*) \cap (S_2=S_2^*)\cap (v_j \in \mathcal{A}_{\ell}) } \Big\}. \label{suba2}
\end{align}

From $(S_1^*,S_2^*) \in \mathcal{L}(K_{1,n},P_n)$, and the definitions of $\mathcal{L}(K_{1,n},P_n)$ and $T(K_{1,n},P_n)$, we have $|S_1^*|=K_{1,n}$, $|S_2^*|=K_{1,n}$ and $S_1^*\cap S_2^* = \emptyset$.
Note that event $\overline{E_{1j}}
\cap  \overline{E_{2j}}$ means that vertex $v_j$ has no edge with any of $v_1$ and $v_2$; i.e., $S_j$ is a subset of $\mathcal{P}_n \setminus (S_1 \cup S_2)$. Under $v_j \in \mathcal{A}_{\ell}$, it holds that $|S_j|=K_{\ell,n}$. Then conditioning on $(S_1=S_1^*) \cap (S_2=S_2^*)\cap (v_j \in \mathcal{A}_{\ell})$, $\overline{E_{1j}}
\cap  \overline{E_{2j}}$ means that $S_j$ is a $K_{\ell,n}$-size subset of $\mathcal{P}_n \setminus (S_1^* \cup S_2^*)$, which has $P_n-2K_{1,n}$ objects. Thus, we have
\begin{align}
\bP{\overline{E_{1j}}
\cap  \overline{E_{2j}} ~ |~ (S_1=S_1^*) \cap (S_2=S_2^*)\cap (v_j \in \mathcal{A}_{\ell}) } = {{{P_n-2K_{1,n}}\choose{K_{\ell,n}}}\over{{P_n}\choose{K_{\ell,n}}}}. \label{suba3}
\end{align}

Substituting (\ref{suba2}) and (\ref{suba3}) into (\ref{suba1}), we derive
\begin{align}
& \bP{  \bigg[\bigcap_{j=3}^{n} \big(\hspace{1.5pt}  \overline{E_{1j}}
\cap  \overline{E_{2j}} \hspace{1.5pt} \big) \bigg]~|~ (v_1\in \mathcal{A}_1) \cap (v_2\in \mathcal{A}_1) \cap  \overline{E_{12}} } \nonumber \\
& \quad =
\left\{\sum_{\ell=1}^{m}\left[a_{\ell} {{{P_n-2K_{1,n}}\choose{K_{\ell,n}}}\over{{P_n}\choose{K_{\ell,n}}}}\right]\right\}^{n-2}
\sum_{(S_1^*,S_2^*) \in \mathcal{L}(K_{1,n},P_n)}\bP{(S_1=S_1^*) \cap (S_2=S_2^*)}\nonumber \\
& \quad = \left\{\sum_{\ell=1}^{m}\left[a_{\ell} {{{P_n-2K_{1,n}}\choose{K_{\ell,n}}}\over{{P_n}\choose{K_{\ell,n}}}}\right]\right\}^{n-2}, \label{suba5}
\end{align}
where the last step applies
$\sum_{(S_1^*,S_2^*) \in \mathcal{L}(K_{1,n},P_n)}\bP{(S_1=S_1^*) \cap (S_2=S_2^*)}=1 $.

Then we use (\ref{suba5}) and (\ref{suba6}) in (\ref{suba7}) to establish
\begin{align}
 \bE{ \psi_{n,1}  \psi_{n,2}  } \leq {a_1}^2 \left\{\sum_{\ell=1}^{m}\left[a_{\ell} {{{P_n-2K_{1,n}}\choose{K_{\ell,n}}}\over{{P_n}\choose{K_{\ell,n}}}}\right]\right\}^{n-2} \label{secondm}
 \end{align}

Then in view of (\ref{firstm})   (\ref{secondm}) and $\lim_{n\to\infty}(1 - b_{1,n})=1$ from (\ref{pntlnnalp}), we derive
\begin{align}
\frac{\bE{ \psi_{n,1}
\psi_{n,2}  }}
     {\left (  \bE{ \psi_{n,1}  } \right )^2 }
  &\leq \frac{{a_1}^2 \left\{\sum_{\ell=1}^{m}\left[a_{\ell} {{{P_n-2K_{1,n}}\choose{K_{\ell,n}}}\over{{P_n}\choose{K_{\ell,n}}}}\right]\right\}^{n-2}}{[a_1 (1- b_{1,n})^{n-1}]^2} %\nonumber \\
%&
  \leq \left\{\frac{\sum_{\ell=1}^{m}\left[a_{\ell} {{{P_n-2K_{1,n}}\choose{K_{\ell,n}}}\over{{P_n}\choose{K_{\ell,n}}}}\right]}{(1- b_{1,n})^2}\right\}^{n-2} \cdot [1+o(1)] .\label{firstoversecondsbbssa}
\end{align}%[1 + o(1)] \times
%Recalling the expression of $b_{1,n}$ in the left hand side of (\ref{eq:scalinglawxaa}), we obtain
%\begin{align}
% \frac{\bE{ \psi_{n,1}
%\psi_{n,2}  }}
%     {\left (  \bE{ \psi_{n,1}  } \right )^2 }
% \leq \left\{\frac{\sum_{\ell=1}^{m}\left[a_{\ell} {{{P_n-2K_{1,n}}\choose{K_{\ell,n}}}\over{{P_n}\choose{K_{\ell,n}}}}\right]}{\left\{\sum_{\ell=1}^{m}\left[a_{\ell} {{{P_n-K_{1,n}}\choose{K_{\ell,n}}}\over{{P_n}\choose{K_{\ell,n}}}}\right]\right\}^2}\right\}^{n-2} \cdot [1+o(1)] .\label{firstoversecond}
%  \end{align}%[1 + o(1)] \times
In Appendix A of the online full version \cite{fullpdf}, we establish
\begin{align}
\left\{\frac{\sum_{\ell=1}^{m}\left[a_{\ell} {{{P_n-2K_{1,n}}\choose{K_{\ell,n}}}\over{{P_n}\choose{K_{\ell,n}}}}\right]}{\left\{\sum_{\ell=1}^{m}\left[a_{\ell} {{{P_n-K_{1,n}}\choose{K_{\ell,n}}}\over{{P_n}\choose{K_{\ell,n}}}}\right]\right\}^2}\right\}^{n-2} \leq [1+o(1)]. \label{firstoversecondsbbss}
  \end{align}
  Recalling the expression of $b_{1,n}$ in the left hand side of (\ref{eq:scalinglawxaa}), we clearly obtain (\ref{eq:ZeroLaw+NodeIsolation+convergence})    from (\ref{firstoversecondsbbssa}) and (\ref{firstoversecondsbbss}).

\section{Establishing Lemma 2}% \ref{lem_Gq_no_isolated_but_not_conn}}
\label{sec:lem_Gq_no_isolated_but_not_conn}

For convenience, we use $F_n$ to denote the event that graph
 $\mathbb{G}(n,\overrightarrow{a},\overrightarrow{K_n},P_n)$ has no isolated vertex, but is not connected.
 The basic idea to prove Lemma \ref{lem_Gq_no_isolated_but_not_conn} is to find an
upper bound on the probability $\bP{F_n}$
and then to demonstrate that this bound converges to zero as $n\to \infty$.

 %This approach is similar to the one used for proving the
%one-law for connectivity in $\mathbb{G}_1$ by Ya\u{g}an \cite{yagan_onoff}.

 We use $\mathcal{N}$ to denote the collection of all non-empty
subsets of the vertex set $\{ v_1, \ldots , v_n \}$.
Similar to Ya\u{g}an \cite{yagan_onoff} and Zhao \emph{et al.} \cite{ZhaoYaganGligor}, we set
\[
r_n^{*}  := \min \left ( \left
\lfloor \frac{P_n}{K_{1,n}} \right \rfloor, ~\left \lfloor \frac{n}{2}
\right \rfloor \right ) .
\]
and
\begin{eqnarray}
X_{n,i}= \left \{
\begin{array}{ll}
\max\{ \left \lfloor (1+\varepsilon) K_n \right \rfloor ,
\left \lfloor \lambda K_{1,n} i \right \rfloor \} & ~ \mbox{for $i=2, \ldots, r_n^{*}$}, \\
\lfloor \mu P_n \rfloor, &~ \mbox{for $i=r_n^{*}+1, \ldots, n$},
\end{array}
\right.  \label{eq:X_S_theta}
\end{eqnarray}
for an arbitrary constant $\varepsilon$ with $0<\varepsilon <1$, and some constants $ \lambda, \mu$ that satisfy   $0<\lambda<\frac{1}{2}$, $0<\mu<\frac{1}{2}$, and are selected to ensure
\cite[Equations (43) and (44)]{yagan_onoff}.

Then with $\boldsymbol{X}_n$ denoting the vector $({X}_{n,1},~{X}_{n,2},~
\ldots,~ {X}_{n,n})$, we define an
event $E_n(\boldsymbol{X}_n)$ through
\begin{equation}\nonumber
E_n(\boldsymbol{X}_n)= \bigcup_{T \subseteq \mathcal{N}: ~
|T| \geq 2} ~ \left[\left|\cup_{i \in T}
S_i\right|~\leq~{X}_{n,|T|}\right].
\label{eq:E_n_defn}
\end{equation}
By a crude bounding argument, we   get
\begin{eqnarray}\nonumber
 \bP{ F_n }
 \leq
\bP{E_n(\boldsymbol{X}_n)} + \bP{ F_n \cap \overline{E_n(\boldsymbol{X}_n)} }.  \label{eq:X_S_thetalim0x}
\end{eqnarray}
From (\ref{eq:X_S_thetalim0x}), a proof of Lemma \ref{lem_Gq_no_isolated_but_not_conn}
reduces to establishing the following two propositions.

\begin{proposition}\label{prop:OneLawAfterReductionPart1xx}
Under the conditions of Theorem \ref{thm:OneLaw+NodeIsolation}, it holds that
\begin{eqnarray}
\lim_{n \rightarrow \infty} \bP{E_n(\boldsymbol{X}_n)} =
0. \label{eq:X_S_thetalim0}
\end{eqnarray}
\end{proposition}

\begin{proposition}
{
Under the conditions of Theorem \ref{thm:OneLaw+NodeIsolation}, it holds that
\begin{equation}\nonumber
 \bP{ F_n \cap \overline{E_n(\boldsymbol{X}_n)} } = o(1).
\label{eq:OneLawAfterReductionPart2}
\end{equation}} \label{prop:OneLawAfterReductionPart2}
\end{proposition}

The proofs of Propositions \ref{prop:OneLawAfterReductionPart1xx} and \ref{prop:OneLawAfterReductionPart2} are given in   Appendix B and Appendix C of the online full version \cite{fullpdf}, respectively.

\section{Related Work} \label{related}

The general random intersection graph model $\mathbb{G}(n,\overrightarrow{a}, \overrightarrow{K}_n,P_n)$ is first studied by Godehardt and Jaworski \cite{GodehardtJaworski}, who give a result on the absence of isolated vertex. Afterwards,
  Goderhardt \emph{et al.} \cite{GodehardtJaworskiRybarczyk} extend the result to connectivity. However, the results of both work \cite{GodehardtJaworski,GodehardtJaworskiRybarczyk} require $P_n = O\left(\frac{n}{\log n}\right)$, which is not applicable to practical secure sensor networks, in which $P_n$ grows at least linearly with the number of sensors $n$ have reasonable resiliency against sensor capture attacks \cite{DiPietroTissec,yagan_onoff,ZhaoYaganGligor}. As proved by \cite{DiPietroTissec}, $P_n$ needs to be $\Omega\left(n\right)$ so that an adversary capturing $o(n)$ sensors can only compromise an $o(1)$ portion of sensor communications. In addition, Bloznelis \emph{et al.} \cite{Rybarczyk} investigate  component evolution in $\mathbb{G}(n,\overrightarrow{a}, \overrightarrow{K}_n,P_n)$ and present   conditions for the existence of a {\em giant component} (i.e., a connected component of $\Theta(n)$ vertices). Recently, Zhao \emph{et al.} \cite{ZhaoCDC}   consider $k$-connectivity of general random intersection graphs. Later,  Ya\u{g}an \cite{yagan2015zero} shows that the result of Zhao \emph{et al.} \cite{ZhaoCDC} is constrained
to very narrow parameter ranges and is   not applicable to real-world secure sensor networks. Ya\u{g}an \cite{yagan2015zero} establishes a zero-one law of connectivity in $\mathbb{G}(n,\overrightarrow{a}, \overrightarrow{K}_n,P_n)$. Specifically,   recalling that $b_{1,n}$ is the probability that a typical vertex in group $\mathcal{A}_1$ has an edge with another typical vertex in $\mathcal{V}_n$ ($b_{1,n}$ equals the left hand side of (\ref{eq:scalinglawxaa}), we rewrite
Ya\u{g}an's result as follows:
\begin{displayquote}
{\em For a graph $\mathbb{G}(n,\overrightarrow{a}, \overrightarrow{K_n},P_n)$ under $ P_n =
\Omega(n)$ and $\omega\Big(\sqrt{\frac{P_n}{n}}\Big) = K_{1,n} \leq K_{2,n}\leq \ldots \leq K_{m,n} = o\Big(\sqrt{\frac{P_n(\ln n)^2}{n}}\Big)$,}\\\indent{\em if there exists a positive constant $c$ such that
\begin{align}
b_{1,n} & \sim \frac{c\ln  n }{n},   \label{eq:scalinglaw_old_2}
\end{align}}
{\em then it holds that
\begin{align}
\lim_{n \rightarrow \infty }\hspace{-1pt} \mathbb{P}\hspace{-1pt}\bigg[
\hspace{-3pt}\begin{array}{c}
\mathbb{G}(n,\overrightarrow{a}, \overrightarrow{K_n},P_n) \\
\mbox{is connected.}
\end{array}\hspace{-3pt}
\bigg] = \begin{cases} 0,\quad\hspace{-4pt}\text{if  }c<1,\\ 1,\quad\hspace{-4pt}\text{if  }c>1.\end{cases}  \label{eq:scalinglawxaaccster}
\end{align}}
%\begin{subnumcases}
%{ \hspace{-6pt}  \lim_{n \rightarrow \infty }\hspace{-1pt} \mathbb{P}\hspace{-1pt}\bigg[
%\hspace{-3pt}\begin{array}{c}
%\mathbb{G}(n,\overrightarrow{a}, \overrightarrow{K_n},P_n) \\
%\mbox{is connected.}
%\end{array}\hspace{-3pt}
%\bigg] \hspace{-2pt}=\hspace{-2pt}}  \hspace{-3pt}0,\quad\hspace{-4pt}\text{if  }\lim_{n \to \infty}{\beta_n}  \hspace{-1.5pt} =  \hspace{-1.5pt} - \infty, \label{thm-con-eq-0xaacc} \\
%\hspace{-3pt}1,\quad\hspace{-4pt}\text{if  }\lim_{n \to \infty}{\beta_n}    \hspace{-1.5pt}=  \hspace{-1.5pt}  \infty. \label{thm-con-eq-1xaacc}
%\end{subnumcases}
\end{displayquote}

 However, as we explain below, our result outperforms Ya\u{g}an's result \cite{yagan2015zero} in the following two aspects. First, our zero-one law is more fine-grained than that of Ya\u{g}an \cite{yagan2015zero}.
 In a nutshell, with $\beta_n$ given by (\ref{eq:scalinglaw}), the scaling condition
(\ref{eq:scalinglaw_old_2}) enforced by Ya\u{g}an \cite{yagan2015zero}
requires a deviation of $\beta_n = \pm \Omega(\ln n)$
 to get the zero-one law, whereas in our formulation (\ref{eq:scalinglaw}),
it suffices to have an unbounded deviation; e.g., even $\beta_n =  \pm \Theta (\ln n \ln n), \pm \Theta (\ln n \ln n\ln n)$ will do (note that Section 1 already has a discussion on this).
Put differently, we cover the case of $c=1$ in (\ref{eq:scalinglawxaaccster}) under (\ref{eq:scalinglaw_old_2})
and show that $\mathbb{G}(n,\overrightarrow{a}, \overrightarrow{K}_n,P_n)$ could be
 connected or disconnected with high probability, depending on the limit
of $\beta_n$. Second, our condition $\omega(1) = K_{1,n} \leq K_{2,n}\leq \ldots \leq K_{m,n} = o(\sqrt{P_n})$ is broader than the condition by Ya\u{g}an \cite{yagan2015zero}: $ P_n =
\Omega(n)$ and $\omega\big(\sqrt{\frac{P_n}{n}}\big) = K_{1,n} \leq K_{2,n}\leq \ldots \leq K_{m,n} = o\big(\sqrt{\frac{P_n(\ln n)^2}{n}}\big)$, and thus
 has broader applicability in secure sensor networks and social networks. The first point (i.e., a more fine-grained zero-one law) of the two points above is the major improvement of our work over Ya\u{g}an's result \cite{yagan2015zero}. For a few other random graphs, this kind of  improvement is the   focus of several work \cite{ZhaoYaganGligor,han2007very,MakowskiAllerton,ISIT,FJYGISIT2014,YKKKK} as well.

\section{Conclusion} \label{sec:Conclusion}

In this paper, we derive a sharp zero-one law for connectivity in a general random intersection graph. Our result can be applied to secure sensor networks and social networks.
 A general random intersection graph  is defined on a vertex set $\mathcal {V} = \{v_1,
v_2, \ldots, v_n \}$ as follows. Each vertex $v_i$ ($i=1,2,\ldots,n$)
is assigned an object set $S_i$ from an object pool $\mathcal {P}_n$
comprising $P_n$ distinct objects. Each object set $S_i$ is formed according to the following two-step
procedure: First, the size of $S_i$, $|S_i|$, is determined
according to the following probability distribution $\bP{v \in \mathcal{A}_i}= a_i$, where $\sum_{i=1}^{m}a_i  = 1$. Next, $S_i$ is constructed by
selecting $|S_i|$ distinct objects uniformly at random from the
object pool $\mathcal {P}_n$. This process is repeated independently for all
object sets $S_1, \ldots, S_n$. Finally, an undirected edge is
assigned between two vertices if and only if their corresponding object
sets have at least one object in common; namely, distinct vertices
$v_i$ and $v_j$ have an edge in between if and only if $S_i \bcap
S_j \neq \emptyset$.

 \small

\bibliographystyle{abbrv}

\end{document}

\end{document}